\documentclass[12 pt, amsfonts,amsmath, amssymb,color]{article}
\usepackage{authblk}
\def\baselinestretch{1.0}
\usepackage{amsmath,amssymb,amsfonts,latexsym,float,graphics,epsfig}
\usepackage{subfig}
\usepackage{verbatim}
\usepackage{relsize}
\usepackage{hyperref}
\usepackage{graphicx}
\usepackage{bm}
\usepackage{epstopdf}
\usepackage{slashed}
\usepackage{color}
\usepackage{tikz-cd}
\usepackage[utf8]{inputenc}
\usepackage[T1]{fontenc}

\def\be{\begin{equation}}
\def\ee{\end{equation}}
\def\bea{\begin{eqnarray}}
\def\eea{\end{eqnarray}}

\begin{document}
%<<<<<<<<<<< enumeration of eqns section wise>>>>>>>>>>>>>>>>>>>

\renewcommand\theequation{\arabic{section}.\arabic{equation}}
\catcode`@=11 \@addtoreset{equation}{section}
%<<<<<<<<<<<<<<<<<<<<<<<<<<<<<<<<<>>>>>>>>>>>>>>>>>>>>>>>>>>>>>>>>>
\newtheorem{axiom}{Definition}[section]
\newtheorem{theorem}{Theorem}[section]
\newtheorem{axiom2}{Example}[section]
\newtheorem{lem}{Lemma}[section]
\newtheorem{prop}{Proposition}[section]
\newtheorem{cor}{Corollary}[section]

\newcommand{\ben}{\begin{equation*}}
\newcommand{\een}{\end{equation*}}

%%%%%%%%%%%
\let\endtitlepage\relax

\begin{titlepage}
\begin{center}
\renewcommand{\baselinestretch}{1.5}  %Line spacing
%\setstretch{1.5}

\vspace*{-0.5cm}

{\Large {Generalized Li\'enard systems and isochronous connections}}

\vspace{5mm}
\renewcommand{\baselinestretch}{1}  %Line spacing
%\setstretch{1}

\centerline{{\bf Bijan Bagchi$^*$\footnote{bbagchi123@gmail.com}, A. Ghose-Choudhury$^\dagger$\footnote{aghosechoudhury@gmail.com}, Aritra Ghosh$^\ddagger$\footnote{ag34@iitbbs.ac.in, aritraghosh500@gmail.com}, Partha Guha$^\#$\footnote{partha.guha@ku.ac.ae}}}

\vspace{3mm}
\normalsize
\text{$^*$Department of Applied Mathematics, University of Calcutta,}\\
\text{Kolkata, West Bengal 700009, India}\\
\vspace{1.5mm}
\text{$^\dagger$Department of Physics, Diamond Harbour Women’s University,}\\
\text{D.H. Road, Sarisha, West Bengal 743368, India}\\
\vspace{1.5mm}
\text{$^\ddagger$School of Basic Sciences, Indian Institute of Technology Bhubaneswar,}\\
\text{Jatni, Khurda, Odisha 752050, India}\\
\vspace{1.5mm}
\text{$^\#$Department of Mathematics, Khalifa University of Science and Technology,}\\
\text{Main Campus, P.O. Box -127788, Abu Dhabi, United Arab Emirates}\\
\vspace{5mm}

%--------------------------------------------------------------------------------------------------------------------------------------
\begin{abstract}
In this paper, we explore some classical and quantum aspects of the nonlinear Li\'enard equation $\ddot{x} + k x \dot{x} + \omega^2 x + (k^2/9) x^3 = 0$, where $x=x(t)$ is a real variable and $k, \omega \in \mathbb{R}$. We demonstrate that such an equation could be derived from an equation of the Levinson-Smith kind which is of the form $\ddot{z} + J(z) \dot{z}^2 + F(z) \dot{z} + G(z) = 0$, where $z=z(t)$ is a real variable and $\{J(z), F(z), G(z)\}$ are suitable functions to be specified. It can further be mapped to the harmonic oscillator by making use of a nonlocal transformation, establishing its isochronicity. Computations employing the Jacobi last multiplier reveal that the system exhibits a bi-Hamiltonian character, i.e., there are two distinct types of Hamiltonians describing the system. For each of these, we perform a canonical quantization in the momentum representation and explore the possibility of bound states. While one of the Hamiltonians is seen to exhibit an equispaced spectrum with an infinite tower of states, the other one exhibits branching but can be solved exactly in closed form for certain choices of the parameters.
\end{abstract}
\end{center}
\vspace*{0cm}

% \renewcommand{\baselinestretch}{0.2}
% \setlength{\baselineskip}{0.1\baselineskip}
% \setstretch{0.1}

 %\tableofcontents
\end{titlepage}
%\vspace*{0cm}
%%%%%%%5
%\clearpage

\section{Introduction}
Li\'{e}nard equations are a class of nonlinear (generally dissipative) and second-order differential equations \cite{lie1, lie2} initially proposed to address the problem of a damped pendulum. For the real-valued variable \(x\), a Li\'{e}nard equation reads
\begin{equation}\label{1}
    \ddot{x} + f(x) \dot{x} + g(x) = 0,
\end{equation}
where the underlying functions $f(x)$ and $g(x)$ are suitable well-behaved functions (often but not always polynomials). Such systems are of relevance in electrical oscillations \cite{vdP}, optics \cite{dor}, etc. While the non-invariance of a Li\'{e}nard equation under time-reversal indicates that the dynamics is generally dissipative, if \(f(x)\) has zeroes, it opens up the possibility of encountering limit cycles on the phase space if one views the Li\'enard equation as a planar system by identifying $\dot{x} = y, \quad \dot{y} = -g(x) - f(x) y$ \cite{vill1}. Notably, an intriguing case arises for \(f(x) = kx\) and \(g(x) = \omega^2 x + (k^2/9) x^3\) \cite{senti2}, i.e., 
\begin{equation}\label{2}
    \ddot{x} + kx\dot{x} + \omega^2 x + \frac{k^2}{9} x^3 = 0, \quad \quad k, \omega \in \mathbb{R}.
\end{equation} 
 In this case, the nonlinear oscillator (\ref{2}) admits the following periodic solution (see for example, \cite{rosu}):
\begin{equation}\label{3}
    x(t) = \frac{A_0\sin (\omega t + \theta)}{1-\frac{kA_0}{3\theta} \cos (\omega t + \theta)},
\end{equation}
where $A_0$ and $\theta$ are integration constants with the restricted domain $-1<\frac{kA_0}{3\theta}<1$. This indicates an intimate relationship between the Li\'enard system (\ref{2}) and the harmonic oscillator with both the systems supporting isochronous oscillations, i.e., oscillations for which the periodicity is independent of the energy and depends only upon the constant \(\omega\), independent of \(k\). The observed correspondence between isochronous systems and equally-spaced quantum spectra \cite{isospec} suggests that the Li\'{e}nard system (\ref{2}) may admit a harmonic-oscillator-like spectrum \cite{bag1} (see also, \cite{bagconf,senti1}). The important point here is that due to the nonstandard form of the Hamiltonian function, the `quantum mechanics' is worked out in the momentum representation in which one also encounters the notion of a momentum-dependent mass \cite{bag1,bagconf,senti1,bag3}.\\

The purpose of the present investigation is to have a relook at the Li\'enard system (\ref{2}) in order to highlight some intriguing facets of the problem. We will first demonstrate that generic Li\'enard systems (\ref{1}) can arise from certain generalized equations of the Levinson-Smith type \cite{LSE,LSE1}. For such Levinson-Smith equations, we will assess the feasibility of a nonlocal transformation that maps the dynamics to that of the harmonic oscillator; as a result, it will be shown that the system (\ref{2}) can indeed be mapped to the harmonic oscillator, thus proving the isochronicity of the dynamics. Then, we will demonstrate the existence of a bi-Hamiltonian structure in the sense that there are two distinct classes of Hamiltonians which can describe the system consistently. Such Hamiltonians would turn out to be of the nonstandard kind in which there is no direct identification of kinetic and potential energies, although they admit functional forms in which the roles of the coordinate and the momentum appear interchanged with the appearance of momentum-dependent masses and potentials \cite{bag1,bag3}. For the derivation of such aspects, we will follow the preceding works \cite{whitt,nucci1,nucci2,agpgk,pgcar1,pgcar2} which have involved the use of the Jacobi last multiplier to derive Lagrangians (and hence, Hamiltonians) for Li\'enard-type systems. Following this, we will proceed onto studying the quantum aspects of the system by performing a canonical quantization in the momentum representation, following \cite{bag1} and employing the ordering strategy due to von Roos \cite{vonroos} (see \cite{L2quant,L2quant1,VR0,VR1,VR2,VR3,VR4,VR5} for its application in systems with position-dependent masses). The time-independent Schr\"odinger equations for both the cases will be mapped to those of constant-mass systems but with certain effective potentials and the possibility of bound states will be investigated. \\

With this brief introduction, let us present the organization of the paper. In Sec. (\ref{LSMainSec}), we shall describe how Li\'enard-type systems can be obtained from certain Levinson-Smith-type equations by the means of a transformation. Thus, the system (\ref{2}) can be understood as a special case of a Levinson-Smith-type equation which will be discussed in Sec. (\ref{Obtaining}) and the isochronicity condition shall be verified in Sec. (\ref{levinsonsec}). Then, in Sec. (\ref{JLMSec}), we will discuss the Lagrangian and Hamiltonian aspects of the system (\ref{2}), exploring the two classes of Hamiltonians. Following this, in Secs. (\ref{quantizationsec}) and (\ref{boundsec}), canonical quantization in the momentum space will be presented. Our principal conclusions shall then be summarized in Sec. (\ref{concsec}).

\section{Levinson-Smith equations and isochronous oscillations}\label{LSMainSec}
Consider the following family of Levinson-Smith equations for the real-valued variable \(z\):
\begin{equation}\label{LS}
\ddot{z} + J(z) \dot{z}^2 + F(z) \dot{z} + G(z) = 0,
\end{equation} where \(J(z)\), \(F(z)\), and \(G(z)\) are suitable functions that are differentiable over a certain domain (subset of \(\mathbb{R}\)) on which the problem is defined but are not-necessarily polynomials. In \cite{LSE}, the Lagrangian and Hamiltonian functions were obtained by employing the Jacobi last multiplier when a certain condition was satisfied. Now, let us perform a change of variable as dictated by
\begin{equation}\label{mapLStoL}
x = \int \exp \bigg( \int^z J(s) ds \bigg) dz,
\end{equation}
provided \(J(z)\) is of the form of a logarithmic derivative, i.e., \(J(z) = \frac{d(\ln \Phi(z))}{dz}\). Thus, \(x = \int \Phi(s) ds\) and further, assuming the invertibility of this transformation, it follows that (\ref{LS}) can be reduced to the Li\'enard equation
\begin{equation}
\ddot{x} + F(z(x)) \dot{x} + \Phi(z(x)) G(z(x)) = 0.
\end{equation}
Relabeling \(f(x) \equiv F(z(x))\) and \(g(x) \equiv \Phi(z(x)) G(z(x))\) enables us to obtain (\ref{1}). \\

\subsection{A special case}\label{Obtaining}
In order to obtain (\ref{2}), let us take
\begin{equation}\label{exampleLS}
J(z) = -\frac{2}{z}, \quad \quad F(z) = - \frac{\alpha_1}{z}, \quad \quad G(z) = - \frac{c_{-1}}{z} - c_1 z,
\end{equation} where \((\alpha_1,c_{-1}, c_1)\) are real constants. Thus, we can write \(\Phi(z) = \frac{\alpha_2}{z^2}\), where \(\alpha_2\) is some real constant and which directly implies that
\begin{equation}\label{ztoxexample}
x = - \frac{\alpha_2}{z}, \quad \quad F(z(x)) = \frac{\alpha_1 x}{\alpha_2} , \quad \quad G(z(x)) = \frac{c_{-1}x}{\alpha_2}  + \frac{c_1 \alpha_2}{x}.
\end{equation}
It is then immediately implied that one gets the Li\'enard equation (\ref{1}) for the functions
\begin{equation}\label{fxchoicespecialcase}
f(x) = \bigg(\frac{\alpha_1}{\alpha_2}\bigg) x, \quad \quad g(x) = \bigg( \frac{c_{-1}}{\alpha_2^2}\bigg) x^3 + c_1 x.
\end{equation}
Defining \( \alpha_1/\alpha_2 = k\), \(c_1 = \omega^2\), and \(c_{-1}/\alpha_2^2 = k^2/9\), we can obtain our system of interest which is (\ref{2}). 

\subsection{Map to harmonic oscillator}\label{levinsonsec} 
The isochronicity condition for Li\'enard-type systems is well known \cite{sab,isolie}. We will now introduce a nonlocal transformation to map the system (\ref{LS}) to a harmonic oscillator \(\frac{d^2X}{dT^2} + \Omega^2 X = 0\) for constant \(\Omega\) (thus \(X\) evolves with constant period, independent of energy) \cite{isoagcpg}:
\begin{equation}
\frac{dX}{X} = A(z) dz + B(z) dt, \quad \quad T = t,
\end{equation}  where the functions \(A(z)\) and \(B(z)\) are to be determined. This immediately gives
\begin{equation}
\frac{\dot{X}}{X} = A(z) \dot{z} + B(z), 
\end{equation} where the `dot' indicates derivative with respect to \(T\) which coincides with that with respect to \(t\) since \(T = t\). A further differentiation with respect to \(T\) gives
\begin{equation}
\ddot{X} = X \bigg[ \bigg( \frac{dA(z)}{dz} + A(z)^2 - A(z) J(z) \bigg)\dot{z}^2 + \bigg( \frac{d B(z)}{d z} + 2 A(z) B(z) - A(z) F(z) \bigg) \dot{z} + \big(B(z)^2 - A(z) G(z)\big) \bigg],
\end{equation} where we have made use of (\ref{LS}). In order to obtain the harmonic oscillator, one must set
\begin{equation}\label{isocond1}
\frac{dA(z)}{dz} + A(z)^2 - A(z) J(z) = 0, \quad \frac{dB(z)}{d z} + 2 A(z) B(z) - A(z) F(z) = 0, \quad B(z)^2 - A(z) G(z) = - \Omega^2,
\end{equation} where \(\Omega^2 > 0\). As our interest is in the Li\'enard system (\ref{2}), following (\ref{exampleLS}), let us take \(J(z) = -2/z\) which means the first equation above gives \(A(z) = -1/z\). Then, the second equation above gives \(B(z) = - \alpha_1/3z\). Plugging this into the third condition gives us
\begin{equation}
G(z) = - \Omega^2 z \bigg(1 + \frac{\alpha_1^2}{9 \Omega^2 z^2} \bigg). 
\end{equation}
A direct comparison with (\ref{exampleLS}) reveals that the choice of \(G(z)\) taken there coincides with the one found above if we set 
\begin{equation}
c_1 = \Omega^2, \quad \quad \frac{c_{-1}}{c_1} = \frac{\alpha_1^2}{9 \Omega^2}.
\end{equation}
These precisely coincide with the choices of the parameters (with \(\Omega^2 = \omega^2\)) made below equation (\ref{fxchoicespecialcase}) in order to recover the Li\'enard system (\ref{2}) from the Levinson-Smith equation (\ref{LS}). Thus, we have established the isochronicity of the system of our interest using the construction of nonlocal transformations applied to Levinson-Smith-type equations.

\section{Lagrangian and Hamiltonian framework}\label{JLMSec}
In this section, we will make use of the Jacobi last multiplier to derive Lagrangian (and subsequently, Hamiltonian) functions appropriate for the Li\'enard system (\ref{2}). The reader is referred to \cite{bag3,whitt,nucci1,nucci2,agpgk,pgcar1,pgcar2} for the technical details.

\subsection{Jacobi last multiplier}
The Jacobi last multiplier \(M\) for a dynamical system is a factor that satisfies the following equation:
\begin{equation}
\frac{d}{dt}(\ln M) + {\rm div} (X) = 0, 
\end{equation} where \(X\) is the dynamical vector field whose integral curves are the phase trajectories. Here, \({\rm div} (X)\) is the usual divergence, evaluated in a local system of coordinates (if the phase space has real dimension \(n\)) \((x_1,x_2, \cdots, x_n)\) and defined in the sense that \(\pounds_X \Omega = ({\rm div} (X))\Omega\), where \(\Omega = dx_1 \wedge dx_2 \wedge \cdots \wedge dx_n\) is the volume-form and \(\pounds_X\) denotes the Lie derivative with respect to \(X\). \\

For our purposes, let us look at evolution equations of the form of an autonomous Newtonian form, i.e., 
\begin{equation}\label{5}
    \ddot{x}=\mathcal{F}(x,\dot{x}),
\end{equation} 
where $\mathcal{F}(x,\dot{x})$ is some well-behaved function of the coordinate \(x\) and the velocity \(\dot{x}\). This can be cast in the first-order form as \(\dot{x} = y\) and \(\dot{y} = \mathcal{F}(x,y)\), indicating that on the two-dimensional phase space, the Jacobi last multiplier satisfies the following equation: 
\begin{equation}\label{6}
    \frac{d}{dt}(\ln M)+\frac{\partial
\mathcal{F}(x,y)}{\partial y} + \frac{\partial y}{\partial x}=0,
\end{equation} where the last term on the left-hand side is obviously zero. It turns out that for a second-order differential equation such as (\ref{5}), the Jacobi last multiplier is directly related to the corresponding Lagrangian (should one exist) by the following simple formula \cite{whitt} (see appendix of \cite{bag3} for some details):
\begin{equation}\label{7}
    M=\frac{\partial^2L}{\partial \dot{x}^2}.
\end{equation}
Thus, the knowledge of the Jacobi last multiplier may allow one to find the corresponding Lagrangian function such that (\ref{5}) is obtained as an Euler-Lagrange equation. 

  \subsection{Lagrangian and Hamiltonian functions}
For the general Li\'enard system (\ref{1}) written in terms of generic functions \(f(x)\) and \(g(x)\), one can formally solve (\ref{6}) to find that the last multiplier is given by
\begin{equation}\label{8}
    M=\exp\left(\int f(x)
dt\right)\equiv u^{1/\ell},
\end{equation} where  $u$ is some nonlocal variable and $\ell$ is a suitable parameter. It is widely recognized that the Li\'{e}nard equation (\ref{1}) can be written
  as the following coupled system of first-order equations:
  \begin{equation}\label{9}
      \dot{u}=\ell uf(x),\quad \quad \dot{x}=u+W(x),
  \end{equation} in which
  $W(x)=\ell^{-1}(g(x)/f(x))$ while the parameter $\ell$ is
  determined by the following condition:
  \begin{equation}\label{10}
      \frac{d}{dx}\left(\frac{g(x)}{f(x)}\right)
  + \ell(\ell+1) f(x) = 0.
  \end{equation}
 Let us note that not all choices for \(f(x)\) and \(g(x)\) would satisfy the condition\footnote{For example, the van der Pol oscillator \cite{vdP}, i.e., \(\ddot{x} + \sigma (1 - x^2) \dot{x} + \Omega_0^2 x = 0\) for \(\sigma, \Omega_0 > 0\) does not satisfy (\ref{10}).} (\ref{10}), often dubbed the Chiellini integrability condition in the context of Abel equations of the first kind \cite{abel}. Thus, for Li\'enard systems for which the functions \(f(x)\) and \(g(x)\) satisfy the Chiellini condition, (\ref{1}) may be rewritten as (\ref{9}). If that is the case, then combining (\ref{8}) with (\ref{9}) gives us the following expression for the Jacobi last multiplier: 
  \begin{equation}
  M = (\dot{x} - W(x))^{1/\ell}.
  \end{equation}
  Integrating with respect to \(\dot{x}\) twice, one finds from (\ref{7}), the result
  \begin{equation}
  L = \frac{\ell^2\left (\dot{x} -\frac{g(x)}{\ell f(x)} \right)^{\frac{2\ell + 1}{\ell}}}{\left (\ell + 1\right ) \left (2 \ell + 1\right )} + \mathcal{A}(x, t) \dot{x} + \mathcal{B}(x, t),
  \end{equation} where the terms \(\mathcal{A}(x,t) \dot{x}\) and \(\mathcal{B}(x,t)\) can be rejected as they can be combined into a total derivative which does not affect the Euler-Lagrange equations. Thus, the Lagrangian, a function of \(x\) and \(\dot{x}\), is given by
    \begin{equation}\label{lag}
  L(x,\dot{x}) = \frac{\ell^2\left (\dot{x} -\frac{g(x)}{\ell f(x)} \right)^{\frac{2\ell + 1}{\ell}}}{\left (\ell + 1\right ) \left (2 \ell + 1\right )} . 
  \end{equation}
  This describes the Li\'enard system provided the condition (\ref{10}) is met. To derive the form of the Hamiltonian corresponding to the Lagrangian (\ref{lag}), let us employ the usual Legendre transformation defining the Hamiltonian to be $H= p \dot{x} - L$, where $p = \frac{\partial L}{\partial \dot{x}}$ is the canonical momentum which reads
\begin{equation}\label{pcanonicalgeneral}
p = \bigg( \frac{\ell}{\ell + 1} \bigg) \left(\dot{x} -\frac{g(x)}{\ell f(x)} \right)^{\frac{\ell + 1}{\ell}}.
\end{equation}
In order to find the Hamiltonian via a Legendre transform, one ought to solve for the velocity \(\dot{x}\) as a function of \(x\) and \(p\) by inverting the above equation. One is then immediately led to a subtlety for possible cases where \(\frac{\ell}{\ell + 1} \notin \mathbb{Z} \), for this would lead to the appearance of multiple roots while solving for \(\dot{x}\). Ignoring this subtlety for the moment, solving (\ref{pcanonicalgeneral}) in favor of \(\dot{x}\) gives us
\begin{equation}\label{15}
\dot{x} = \frac{g(x)}{\ell f(x)}  + \left [\left (\frac{\ell +1}{\ell}\right )p \right]^{\frac{\ell}{\ell + 1}},
\end{equation}
and which implies that $H$ turns out to be
\begin{equation}\label{16}
H (x, p)|_\ell = \frac{g(x)}{\ell f(x)} p  + \frac{\ell}{2 \ell + 1} \bigg( \frac{\ell + 1}{\ell} p \bigg)^{\frac{2\ell + 1}{\ell + 1}}.
\end{equation}
Some simplification is achieved if we consider rescaling of the momentum to define \(\tilde{p} = \frac{\ell + 1}{\ell} p\), enabling us to rewrite the Hamiltonian as
\begin{equation}\label{17}
H (x, \tilde{p})|_\ell =  \frac{g(x)}{(\ell + 1) f(x)} \tilde{p} + \frac{\ell}{2 \ell + 1} \tilde{p} ^{\frac{2\ell + 1}{\ell + 1}}.
\end{equation}
The Poisson bracket gets modified to \(\{x,\tilde{p}\} = \frac{\ell + 1}{\ell}\), which is equivalent to \(\{x,p\} = 1\). Notice that (\ref{17}) can be interpreted as a deformation of the Berry-Keating Hamiltonian \cite{bk}. The connection of the latter with the Riemann zeros was recently probed in \cite{bk1,bk2}.

\subsection{Hamiltonians corresponding to the system of our interest}\label{Hamsec}
    Considering (\ref{2}), one finds after plugging the forms of \(f(x)\) and \(g(x)\) into (\ref{10}) that the Chiellini condition can be satisfied for\footnote{For a plausible generalization, see Appendix (\ref{app0}).}
  \begin{equation}\label{11}
\ell = -\frac{1}{3}, \quad \ell = -\frac{2}{3}. 
\end{equation}
  It is at once apparent that the above Lagrangian (\ref{lag}) is of the nonstandard form in that it involves a complicated expression of the derivative rather than being controlled by the usual kinetic-energy term which is quadratic in velocity. Of late, such non-conventional forms of Lagrangians have been a subject of active pursuance in the literature (see \cite{bag3} for a recent review). These Lagrangians often lead to certain exotic Hamiltonians, as typified by the so-called \textit{branched Hamiltonians} \cite{wilc,curt}, whose connections to problems of nonlinear dynamics, as governed by autonomous differential equations, have been well researched \cite{bag4}. We will distinguish the two solutions of $\ell$ furnished in (\ref{11}) as `class-I' and `class-II' solutions, respectively, for \(\ell = -1/3, -2/3\). 
  
  \subsubsection{Class-I Hamiltonians: Branched pairs}
  For \(\ell = -1/3\), the Hamiltonian (\ref{17}) turns out to be
  \begin{equation}\label{Hclass1}
H_{\rm I}(x,\tilde{p}) \equiv H(x,\tilde{p})|_{\ell = -1/3} = \frac{3 \tilde{p}}{2} \bigg[ \frac{kx^2}{9} + \frac{\omega^2}{k}\bigg] - \sqrt{\tilde{p}}.
\end{equation}
Notice from (\ref{pcanonicalgeneral}) that \(p < 0\) while \(\tilde{p} = - 2p > 0\). Let us note, however, that \(\frac{\ell}{\ell + 1} = - \frac{1}{2} \notin \mathbb{Z}\). This means, while solving (\ref{pcanonicalgeneral}) to yield (\ref{15}), one faces the appearance of a square root. Consequently, for a given value of \(p\) (or \(\tilde{p}\)), one finds two different values of \(\dot{x}\); explicitly, they read
\begin{equation}
\dot{x}^\pm(x,\tilde{p}) = \pm \frac{1}{\sqrt{\tilde{p}}} - \frac{kx^2}{3} - \frac{3 \omega^2}{k},
\end{equation} indicating the appearance of two `branches', denoted by \(\pm\). On one hand, taking the `\(+\)' sign and performing the Legendre transform, one finds the Hamiltonian (\ref{Hclass1}) which we will denote hereafter as \(H^+_{\rm I}(x,\tilde{p})\). On the other hand, taking the `\(-\)' sign gives us the following Hamiltonian: 
\begin{equation}
H^-_{\rm I}(x,\tilde{p}) = \frac{3 \tilde{p}}{2} \bigg[ \frac{kx^2}{9} + \frac{\omega^2}{k}\bigg] + \sqrt{\tilde{p}}. 
\end{equation}
Thus, the case with \(\ell = -1/3\) leads us to a pair of Hamiltonians, the so-called branched partners which can be expressed together as
\begin{equation}\label{classIbranchedH}
H^\pm_{\rm I}(x,\tilde{p}) = \frac{3 \tilde{p}}{2} \bigg[ \frac{kx^2}{9} + \frac{\omega^2}{k}\bigg] \mp \sqrt{\tilde{p}}. 
\end{equation}

  \subsubsection{Class-II Hamiltonian}
For \(\ell = -2/3\), \(\frac{\ell}{\ell + 1} = - 2 \in \mathbb{Z}\). Thus, one does not encounter branching while solving for \(\dot{x}\) as a function of \(p\) or \(\tilde{p}\). The Hamiltonian in this case can be found by putting \(\ell = -2/3\) in (\ref{17}), i.e., 
\begin{equation}\label{Hclass2}
H_{\rm II}(x,\tilde{p}) \equiv H(x,\tilde{p})|_{\ell = -2/3} =  3 \tilde{p} \bigg[ \frac{kx^2}{9} + \frac{\omega^2}{k}\bigg] + \frac{2}{\tilde{p}}.
\end{equation}
Notice from (\ref{pcanonicalgeneral}) that \(p < 0\), while \(\tilde{p} = -p/2 > 0\). Therefore, one finds two classes of Hamiltonians, namely, the class-I and class-II cases, while further, one finds a branched pair of Hamiltonians for the class-I case. 

\subsubsection{Momentum-dependent masses and potentials}
\begin{figure}
\begin{center}
\includegraphics[scale=1.15]{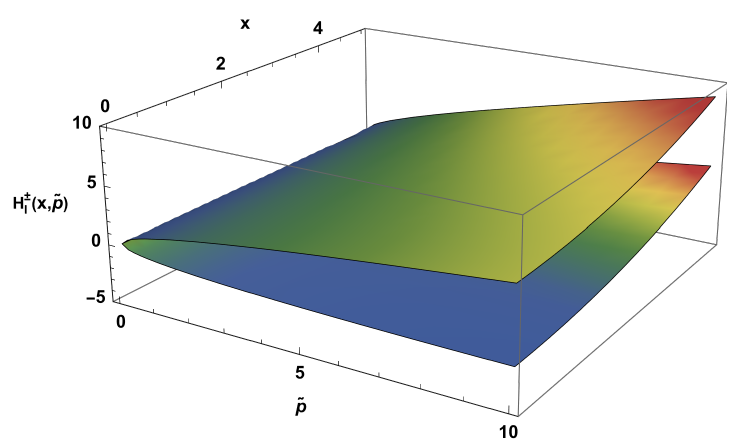}
\caption{3D plot of the branched Hamiltonians \(H^\pm_{\rm I}(x,\tilde{p})\). We have taken \(k = 0.1\) and \(\omega = 0.1\). The branches coalesce at \(\tilde{p} = 0\).}
\label{fig1}
\end{center}
\end{figure}

\begin{figure}
\begin{center}
\includegraphics[scale=1.15]{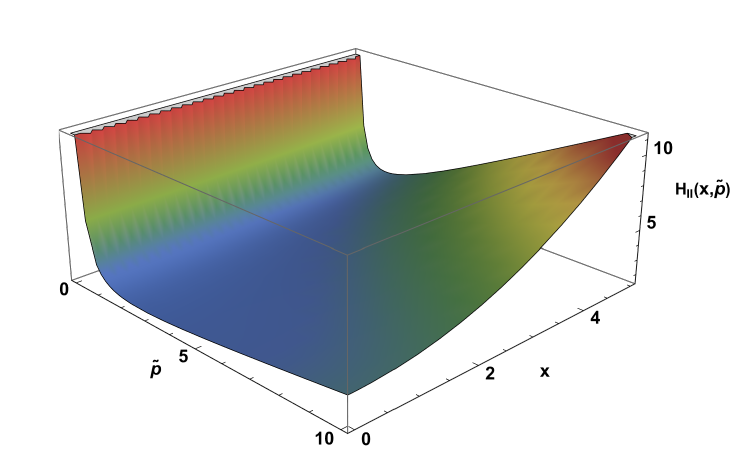}
\caption{3D plot of the Hamiltonian \(H_{\rm II}(x,\tilde{p})\). We have taken \(k = 0.1\) and \(\omega = 0.1\). The singularity at \(\tilde{p}=0\) is clear.}
\label{fig2}
\end{center}
\end{figure}

Interestingly, both classes of Hamiltonians presented above can be cast in the form
\begin{equation}
H(x,\tilde{p}) = \frac{x^2}{2 m(\tilde{p})} + V(\tilde{p}),
\end{equation} where the momentum-dependent masses and potentials are
\begin{equation}\label{massprofiles}
m_{\rm I}(\tilde{p}) = \frac{3}{k\tilde{p}}, \quad \quad m_{\rm II}(\tilde{p}) = \frac{3}{2k\tilde{p}},
\end{equation}
\begin{equation}\label{potprofiles}
V^+_{\rm I}(\tilde{p}) = \bigg( \frac{3 \omega^2}{2k}\bigg) \tilde{p} - \sqrt{\tilde{p}},  \quad \quad V^-_{\rm I}(\tilde{p}) = \bigg( \frac{3 \omega^2}{2k}\bigg) \tilde{p} + \sqrt{\tilde{p}}, \quad \quad V_{\rm II}(\tilde{p}) = \bigg( \frac{3 \omega^2}{k}\bigg) \tilde{p} + \frac{2}{\tilde{p}}.
\end{equation}
Thus, it appears as if the roles of the coordinate and the momentum have become interchanged with the concomitant appearance of momentum-dependent masses and potentials. Note that from (\ref{pcanonicalgeneral}), the whole problem is defined on the half-line of the momentum variable, i.e., \(0<\tilde{p} < \infty\). The concept of a variable mass 
 has found widespread attention in the literature, particularly in the context of position-dependent mass \cite{vonroos,L2quant,L2quant1,VR0,VR1,VR2,VR3,VR4,VR5,bagchi2,mus1,carinena,cruz,que,koc,cunha,mus2,dha,fer,bag5,bpm}. For some early literature, let us refer to the works on compositionally-graded crystals \cite{Gel},
quantum dots \cite{Ser}, and Helium clusters \cite{Bar}. In parallel, the issue of the momentum-dependent-mass counterpart has been studied for classical problems seeking plausible quantum interpretation \cite{bag1} as well as for branched-Hamiltonian models \cite{bag3, curt}. The Hamiltonians have been plotted in Figs. (\ref{fig1}) and (\ref{fig2}). 

\section{Quantization \`a la von Roos: Effective potentials}\label{quantizationsec}
Let us now perform a canonical quantization of the nonlinear Li\'enard system (\ref{2}). Although the forms of the Hamiltonians (\ref{classIbranchedH}) and (\ref{Hclass2}) suggest that quantization is not straightforward in the coordinate representation, it can be performed directly in the momentum representation for which (we will use the units in which \(\hbar = 1\))
\begin{equation}
p \rightarrow \hat{p} = p, \quad \quad x \rightarrow \hat{x} = i \frac{d}{d p},
\end{equation} therefore leading to \([\hat{x},\hat{p}] = i \). In terms of the rescaled momentum \(\tilde{p}\), canonical quantization is achieved as
\begin{equation}\label{canquanttilde}
\tilde{p} \rightarrow \hat{\tilde{p}} = \tilde{p}, \quad \quad x \rightarrow \hat{x} = i \bigg(\frac{\ell + 1}{\ell}\bigg) \frac{d}{d \tilde{p}}.
\end{equation} 
Now, due to the ordering ambiguity of the term involving the momentum-dependent mass in (\ref{classIbranchedH}) and (\ref{Hclass2}) (because position and momentum do not commute), we will use the von Roos stratagem \cite{vonroos} towards obtaining time-independent Schr\"odinger equations whose solutions will provide us with the stationary states. The ordered Hamiltonian operator takes the following form \cite{bag1}: 
\begin{eqnarray}
\hat{H}(\hat{x},\hat{\tilde{p}}) = \frac{1}{4} \bigg[ m^\alpha(\hat{\tilde{p}}) \hat{x} m^\beta (\hat{\tilde{p}}) \hat{x} m^\gamma (\hat{\tilde{p}}) + m^\gamma(\hat{\tilde{p}}) \hat{x} m^\beta (\hat{\tilde{p}}) \hat{x} m^\alpha (\hat{\tilde{p}}) \bigg] + V(\hat{\tilde{p}}) , \label{vonroos}
\end{eqnarray} where the `ambiguity parameters' \(\alpha\), \(\beta\), and \(\gamma\) satisfy \(\alpha + \beta + \gamma = -1\). Combining (\ref{canquanttilde}) with (\ref{vonroos}), the corresponding time-independent Schr\"odinger equation turns out to be (primes will denote derivatives with respect to \(\tilde{p}\))
\begin{eqnarray}
- \bigg(\frac{\ell + 1}{\ell}\bigg)^2 \frac{1}{2 m(\tilde{p})}  \bigg[\psi'' (\tilde{p}) - \frac{m'(\tilde{p})}{m(\tilde{p})} \psi'(\tilde{p}) + \frac{\beta + 1}{2} \bigg( 2 \frac{m'(\tilde{p})^2}{m(\tilde{p})^2}  - \frac{m''(\tilde{p})}{m(\tilde{p})}\bigg) \psi(\tilde{p})  \nonumber \\
 + \alpha(\alpha + \beta + 1) \frac{m'(\tilde{p})^2}{m(\tilde{p})^2} \psi(\tilde{p}) \bigg] 
+ V(\tilde{p}) \psi(\tilde{p}) = E \psi(\tilde{p}). 
\label{TISEgen}
\end{eqnarray}
If \(m(\tilde{p}) \sim 1/\tilde{p}\) as in (\ref{massprofiles}), we will have 
\begin{equation} 
2 \frac{m'(\tilde{p})^2}{m(\tilde{p})^2}  = \frac{m''(\tilde{p})}{m(\tilde{p})}, \quad \quad \frac{m'(\tilde{p})}{m(\tilde{p})} = - \frac{1}{\tilde{p}}.
\end{equation}
Thus, (\ref{TISEgen}) simplifies to  
\begin{equation}\label{TISEsimp}
- \frac{1}{2 m(\tilde{p})}  \bigg[\psi'' (\tilde{p}) + \frac{\psi'(\tilde{p})}{\tilde{p}} + \frac{\alpha(\alpha + \beta + 1)}{\tilde{p}^2} \psi(\tilde{p}) \bigg] + \tilde{V}(\tilde{p}) \psi(\tilde{p}) = \tilde{E} \psi(\tilde{p}), 
\end{equation} 
where
\begin{equation}
\tilde{V}(\tilde{p}) \equiv \bigg(\frac{\ell + 1}{\ell}\bigg)^{-2} V(\tilde{p}), \quad \tilde{E} \equiv \bigg(\frac{\ell + 1}{\ell}\bigg)^{-2} E. 
\end{equation}
Next, we will show that the momentum-dependent-mass problem can be mapped to a constant-mass problem but with an effective potential that differs from the form of \(\tilde{V}(\tilde{p})\). At this stage, let us focus separately on the class-I and class-II Hamiltonians. The corresponding quantities shall be labeled with subscripts `I' and `II', respectively. 

\subsection{Class-I}
Let us at the moment suppress the superscripts \(\pm\) for the branched pairs of the class-I. For \(m_{\rm I}(\tilde{p}) = 3/k \tilde{p}\) as given in (\ref{massprofiles}), (\ref{TISEsimp}) turns out to be
\begin{equation}\label{TISEsimp1}
\tilde{p} \psi''_{\rm I}(\tilde{p}) + \psi'_{\rm I}(\tilde{p}) + \frac{\alpha(\alpha + \beta + 1)}{\tilde{p}} \psi_{\rm I}(\tilde{p}) + \frac{6}{k} (\tilde{E}_{\rm I} - \tilde{V}_{\rm I}(\tilde{p})) \psi_{\rm I}(\tilde{p}) = 0. 
\end{equation} 
It is convenient to introduce the transformation \(\tilde{p} = \eta \xi^2\), where \(0 < \xi < \infty\) and \(\eta\) is a positive constant. This implies that
\begin{equation}\label{TISE1moresimp}
\frac{d^2 \psi_{\rm I}(\xi)}{d \xi^2} + \frac{1}{\xi} \frac{d\psi_{\rm I}(\xi)}{d\xi} - \frac{\epsilon}{\xi^2} \psi_{\rm I}(\xi) + \frac{24 \eta}{k} (\tilde{E}_{\rm I} - \tilde{V}_{\rm I}(\xi)) \psi_{\rm I}(\xi) = 0,
\end{equation}
where \(\epsilon = - 4 \alpha (\alpha + \beta + 1) = 4 \alpha \gamma\). The term with the first derivative can be eliminated by defining a new function
\begin{equation}\label{linearderiremoval}
\phi_{\rm I}(\xi) = \sqrt{\xi} \psi_{\rm I}(\xi),
\end{equation} and thus allows us to rewrite (\ref{TISE1moresimp}) as 
\begin{equation}\label{TISE1moresimp1}
\frac{d^2 \phi_{\rm I}(\xi)}{d \xi^2}  - \frac{\left(\epsilon - \frac{1}{4}\right)}{\xi^2} \phi_{\rm I}(\xi) + \frac{24 \eta}{k} (\tilde{E}_{\rm I} - \tilde{V}_{\rm I}(\xi)) \phi_{\rm I}(\xi) = 0.
\end{equation}
Therefore, the dynamics has been mapped to that of a constant-mass particle but in an effective potential, i.e., 
\begin{equation}\label{TISEunitmassreduced1}
- \frac{d^2 \phi_{\rm I}(\xi)}{d\xi^2} + V_{\rm I}|_{\rm eff} (\xi) \phi_{\rm I}(\xi) = \mathcal{E}_{\rm I} \phi_{\rm I}(\xi),
\end{equation}
with
\begin{equation}
V_{\rm I}|_{\rm eff} (\xi) = \frac{24 \eta}{k } \tilde{V}_{\rm I}(\xi) + \frac{\left(\epsilon - \frac{1}{4}\right)}{\xi^2}, \quad \quad \mathcal{E}_{\rm I} = \frac{24 \eta}{k}\tilde{E}_{\rm I}.
\end{equation}
In particular, corresponding to the momentum-dependent and branched potentials \(V^\pm_{\rm I}(\tilde{p})\) appearing in (\ref{potprofiles}), the effective potentials are
\begin{equation}
V^\pm_{\rm I}|_{\rm eff} (\xi) = \frac{\omega^2}{64} \xi^2 \mp \sqrt{\frac{k}{6}} \frac{\xi}{8} + \frac{\left(\epsilon - \frac{1}{4}\right)}{\xi^2}, \label{Vieff} 
\end{equation} where we chose \(\eta = k/24\) without loss of generality.

\subsection{Class-II}
Next, let us consider the other Hamiltonian (\ref{Hclass2}) from the bi-Hamiltonian pair. Corresponding to the mass function \(m_{\rm II}(\tilde{p}) = 3/2k\tilde{p}\) in (\ref{massprofiles}), (\ref{TISEsimp}) gives
\begin{equation}\label{TISEsimp12}
\tilde{p} \psi''_{\rm II}(\tilde{p}) + \psi'_{\rm II}(\tilde{p}) + \frac{\alpha(\alpha + \beta + 1)}{\tilde{p}} \psi_{\rm II}(\tilde{p}) + \frac{3}{k} (\tilde{E}_{\rm II} - \tilde{V}_{\rm II}(\tilde{p})) \psi_{\rm II}(\tilde{p}) = 0. 
\end{equation} 
As before, let us introduce the transformation \(\tilde{p} = \rho \zeta^2\), where \(0 < \zeta < \infty\) and \(\rho\) is a positive constant. In terms of the new independent variable \(\zeta\), the time-independent Schr\"odinger equation transforms to
\begin{equation}\label{TISE1moresimp2}
\frac{d^2 \psi_{\rm II}(\zeta)}{d \zeta^2} + \frac{1}{\zeta} \frac{d\psi_{\rm II}(\zeta)}{d\zeta} - \frac{\epsilon}{\zeta^2} \psi_{\rm II}(\zeta) + \frac{12 \rho}{k} (\tilde{E}_{\rm II} - \tilde{V}_{\rm II}(\zeta)) \psi_{\rm II}(\zeta) = 0,
\end{equation}
where \(\epsilon\) is defined in the line below (\ref{TISE1moresimp}). Once again, we will perform the transformation
\begin{equation}\label{scalingclassii}
\phi_{\rm II}(\zeta) = \sqrt{\zeta} \psi_{\rm II}(\zeta),
\end{equation} and which implies that
\begin{equation}\label{TISE1moresimp12}
\frac{d^2 \phi_{\rm II}(\zeta)}{d \zeta^2}  - \frac{\left(\epsilon - \frac{1}{4}\right)}{\zeta^2} \phi_{\rm II}(\zeta) + \frac{12 \rho}{k} (\tilde{E}_{\rm II} - \tilde{V}_{\rm II}(\zeta)) \phi_{\rm II}(\zeta) = 0.
\end{equation}
Thus, the Schr\"odinger equation appears in the form of that arising from a constant-mass scenario; explicitly, one can write
\begin{equation}\label{effTISE2}
- \frac{d^2 \phi_{\rm II}(\zeta)}{d\zeta^2} + V_{\rm II}|_{\rm eff} (\zeta) \phi_{\rm II}(\zeta) = \mathcal{E}_{\rm II} \phi_{\rm II}(\zeta),
\end{equation} where we have defined
\begin{equation}
V_{\rm II}|_{\rm eff} (\zeta) = \frac{12 \rho}{k } \tilde{V}_{\rm II}(\zeta) + \frac{\left(\epsilon - \frac{1}{4}\right)}{\zeta^2}, \quad \quad \mathcal{E}_{\rm II} = \frac{12 \rho}{k}\tilde{E}_{\rm II}.
\end{equation}
Thus, referring to the functional form of \(V_{\rm II}(\tilde{p})\) in (\ref{potprofiles}), the effective potential turns out to be
\begin{equation}\label{IIeffexp}
V_{\rm II}|_{\rm eff} (\zeta) = \omega^2 \zeta^2 + \frac{\left(\epsilon - \frac{1}{4} + \frac{96}{k} \right)}{\zeta^2},
\end{equation} where we chose \(\rho = k/12\) for the sake of concreteness. This effective potential was obtained earlier in \cite{bag1}.

\section{Bound states}\label{boundsec}
Having quantized the Li\'enard system (\ref{2}), we demonstrated the equivalence of the momentum-dependent-mass Schr\"odinger equations for class-I and class-II with constant-mass Schr\"odinger equations (\ref{TISEunitmassreduced1}) and (\ref{effTISE2}) with effective potentials (\ref{Vieff}) and (\ref{IIeffexp}), respectively. We will now explore the possibility of bound states. In particular, referring to the parameter \(\epsilon\) defined below (\ref{TISE1moresimp}), three distinct cases may arise: \(\epsilon < 1/4\), \(\epsilon = 1/4\), and \(\epsilon > 1/4\). Different choices for the ambiguity parameters \((\alpha,\beta,\gamma)\) satisfying the constraint \(\alpha + \beta + \gamma = -1\) shall lead to different values of \(\epsilon\). In particular, the ambiguity parameters as suggested by Mustafa and Mazharimousavi (\(\alpha = \gamma = -1/4\), \(\beta = -1/2\)) \cite{VR5}, albeit for position-dependent-mass systems\footnote{This particular choice of the ambiguity parameters is consistent with the `reliability test' put up by Dutra and Almeida \cite{VR4}.}, conforms to the case with \(\epsilon = 1/4\). Further, the values of ambiguity parameters as suggested by Zhu and Kroemer (\(\alpha = \gamma = -1/2\), \(\beta = 0\)) \cite{VR2} imply \(\epsilon = 1 > 1/4\). Finally, the case \(\epsilon < 1/4\) may occur if one follows the choices of ambiguity parameters given by BenDaniel and Duke (\(\alpha = \gamma = 0\), \(\beta = -1\)) \cite{VR0}, Gora and Williams (\(\alpha = -1\), \(\beta = \gamma = 0\)) \cite{VR1}, or Li and Kuhn (\(\alpha = 0\), \(\beta = \gamma = -1/2\)) \cite{VR3}.

\subsection{Class-II}
\begin{figure}
\begin{center}
\includegraphics[scale=1.2]{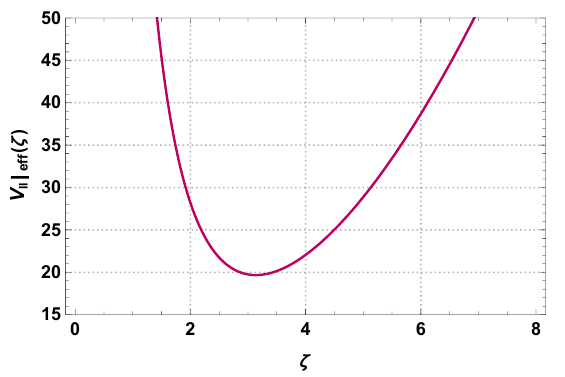}
\caption{Plot of effective potential \(V_{\rm II}|_{\rm eff} (\zeta)\) for \(\epsilon = 1 > 1/4\). We have taken \( k = \omega = 1\) and have observed that for \(\epsilon = 1/4\) or \(0 \leq \epsilon < 1/4\), the general features of the potential remain the same.}
\label{fig4a}
\end{center}
\end{figure}

For this case, the effective potential has been plotted in Fig. (\ref{fig4a}) and clearly shows the isotonic behavior \cite{isotonic,isotonic1,isotonic2}. The wavefunctions can be computed straightforwardly and involve the confluent hypergeometric functions (see Appendix (\ref{appA})) which are in-turn related to the associated Laguerre polynomials \cite{isotonic2,handbook}. The spectrum consists of an infinite tower of states with equispaced energy levels. Defining \(g = \epsilon - \frac{1}{4} +  \frac{96}{k} \), the effective potential \(V_{\rm II}|_{\rm eff}(\zeta)\) reads
\begin{equation}\label{isostandard}
V_{\rm II}|_{\rm eff}(\zeta) = \omega^2 \zeta^2 + \frac{g}{\zeta^2}, \quad \quad \zeta > 0.
\end{equation} Thus, looking at (\ref{effTISE2}), one figures out that the problem is that of a particle with mass \(m = 1/2\) in the isotonic potential. The resulting spectrum turns out to be \cite{isotonic,isotonic1,isotonic2}
\begin{equation}\label{EIIgeneralspectrum}
(E_{\rm II})_n = \frac{1}{4} (\tilde{E}_{\rm II})_n = \omega \bigg(n + \frac{1}{2} + \frac{1}{2} \sqrt{ \epsilon +  \frac{96}{k} }\bigg), \quad \quad n=0,1,2,\cdots. 
\end{equation}
For all values of \(\epsilon \geq 0\) \cite{VR0,VR1,VR2,VR3,VR4,VR5}, the spectrum (level spacing) remains the same although the ground-state energy is dependent on \(\epsilon\) and \(k\). Up to normalization factors, the eigenfunctions of (\ref{effTISE2}) turn out to be (see Appendix (\ref{appA}))
\begin{equation}
\phi_n(\zeta) \sim \zeta^\nu e^{- \frac{\omega \zeta^2}{2}} {_1F_1} \bigg( -n; \nu + \frac{1}{2}; \omega \zeta^2  \bigg), \quad \nu = \frac{1}{2} + \sqrt{\epsilon + \frac{96}{k}},
\end{equation} where \(n = 0,1,2,\cdots\) and \(_1F_1(\cdot;\cdot;\cdot)\) is the confluent hypergeometric function that solves the Kummer differential equation. Notice that the physical wavefunctions are obtained from the above via (\ref{scalingclassii}). 

\subsection{Class-I}\label{classIspectrumsec}
\begin{figure}
\begin{center}
\includegraphics[scale=1.2]{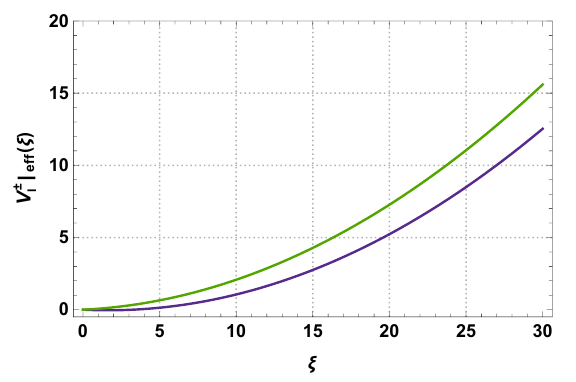}
\caption{Variation of the (branched) effective potentials \(V^\pm_{\rm I}|_{\rm eff} (\xi)\) for \(\epsilon = 1/4\). We have taken \( k = \omega = 1\). Here, \(V^+_{\rm I}|_{\rm eff} (\xi)\) is indicated with violet and \(V^-_{\rm I}|_{\rm eff} (\xi)\) is indicated with green.}
\label{fig4ab1}
\end{center}
\end{figure}

\begin{figure}
\begin{center}
\includegraphics[scale=1.2]{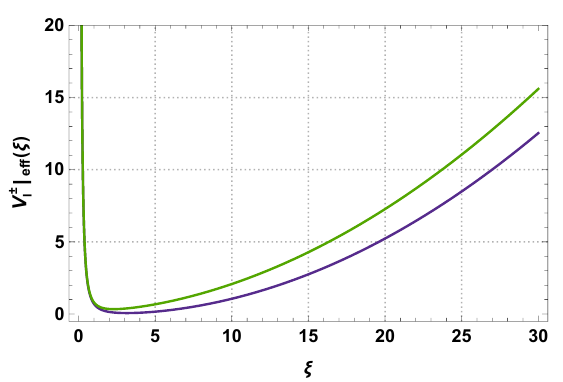}
\caption{Variation of the (branched) effective potentials \(V^\pm_{\rm I}|_{\rm eff} (\xi)\) for \(\epsilon = 1 > 1/4\). We have taken \(k = \omega = 1\). Here, \(V^+_{\rm I}|_{\rm eff} (\xi)\) is indicated with violet and \(V^-_{\rm I}|_{\rm eff} (\xi)\) is indicated with green.}
\label{fig4ab}
\end{center}
\end{figure}

\begin{figure}
\begin{center}
\includegraphics[scale=1.2]{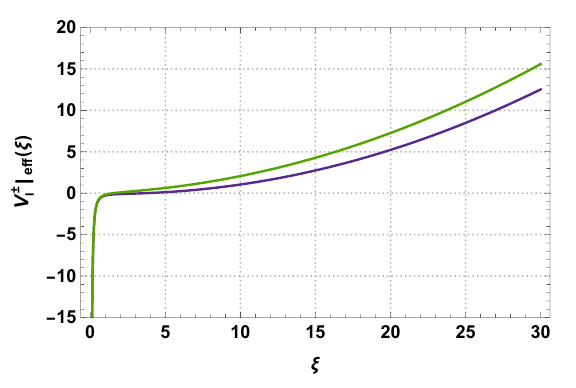}
\caption{Variation of the (branched) effective potentials \(V^\pm_{\rm I}|_{\rm eff} (\xi)\) for \(\epsilon = 0 < 1/4\). We have taken \( k = \omega = 1\). Here, \(V^+_{\rm I}|_{\rm eff} (\xi)\) is indicated with violet and \(V^-_{\rm I}|_{\rm eff} (\xi)\) is indicated with green.}
\label{fig4ab2}
\end{center}
\end{figure}

Let us now come to the other case, namely, the quantum mechanics of the class-I Hamiltonian. With the corresponding effective (branched) potentials given in (\ref{Vieff}), their variation has been depicted in Figs. (\ref{fig4ab1}), (\ref{fig4ab}), and (\ref{fig4ab2}) for the cases \(\epsilon > 1/4\), \(\epsilon = 1/4\), and \(\epsilon < 1/4\), respectively. The effective (branched) potentials can be rewritten as
\begin{equation}\label{Vieffanew}
V^\pm_{\rm I}|_{\rm eff} (\xi) = \frac{\omega^2}{64} (\xi \mp \xi_0)^2 + \frac{\Delta}{\xi^2}  - \frac{k}{24 \omega^2}, \quad \quad \xi_0 = \sqrt{\frac{k}{6}} \frac{4}{\omega^2}, \quad \quad \Delta = \epsilon - \frac{1}{4},
\end{equation}  where \(\xi > 0\). An exact solution in closed form is possible for the special case where \(\epsilon = 1/4\) or \(\Delta = 0\); this may be reached by taking the ordering parameters to be \(\alpha = \gamma = -1/4\), \(\beta = -1/2\), as suggested by Mustafa and Mazharimousavi \cite{VR5}. In this case, one gets
\begin{equation}\label{V1effnew}
V^\pm_{\rm I}|_{\rm eff} (\xi) = \frac{\omega^2}{64} (\xi \mp\xi_0)^2  - \frac{k}{24 \omega^2}, \quad \quad \xi_0 = \sqrt{\frac{k}{6}} \frac{4}{\omega^2}, 
\end{equation} where \(\xi > 0\). While these naively appear to be identical to the harmonic potential (with center at \(\pm \xi_0\)) for a particle of mass \(m = 1/2\) and frequency \(\omega/4 \), there is a peculiarity because \(\xi > 0\) and this must be accounted for. Explicitly, a sufficient condition for the existence of bound states is as follows --

\begin{prop}
A state 
 \begin{equation}\label{hermitesolution}
 \phi^\pm_{{\rm I},n}(\xi) = \chi_n(\xi \mp \xi_0) \sim e^{- \frac{\omega}{16} (\xi \mp \xi_0)^2} H_n\bigg(\sqrt{\frac{\omega}{8}}(\xi \mp \xi_0)\bigg)
 \end{equation}
 is a bound state of the potential (\ref{Vieffanew}) \(V^\pm_{\rm I}|_{\rm eff} (\xi)\) for \(\Delta = 0\), if
  \begin{equation}\label{hermitecondition}
  H_n\bigg(\sqrt{\frac{\omega \xi_0^2}{8}}\bigg) = H_n\bigg(\sqrt{\frac{k}{3 \omega}}\bigg) = 0,
  \end{equation} for some \(n \in \mathbb{N}\). 
  \end{prop}
  
  \textit{Proof --} Since the potential (\ref{Vieffanew}) for \(\epsilon = 1/4\) or \(\Delta = 0\) corresponds to (\ref{V1effnew}) which naively mimics the harmonic-oscillator potential, the equation (\ref{TISEunitmassreduced1}) can be solved in terms of the Hermite polynomials, i.e., one gets solutions corresponding to \(V^\pm_{\rm I}|_{\rm eff} (\xi)\) of the form (\ref{hermitesolution}) with the \(\pm\) signs denoting the two branches. However, because \(\xi > 0\), one also needs to satisfy the boundary condition \( \phi^\pm_{{\rm I},n}(0) = \chi_n(\mp \xi_0) = 0\). Thus, only if the parameters \(k\) and \(\omega\) conform to the condition (\ref{hermitecondition}) does one get a bound state consistent with the boundary conditions.

  \begin{cor}
  If the parameters \((k,\omega)\) of the model are chosen in the manner that one of the branched partners (say, \(V^+_{\rm I}|_{\rm eff} (\xi)\)) admits the bound state \( \phi^+_{{\rm I},n}(\xi) = \chi_n(\xi - \xi_0)\) for some \(n \in \mathbb{N}\), then its branched partner \(V^-_{\rm I}|_{\rm eff} (\xi)\) admits the bound state \( \phi^-_{{\rm I},n}(\xi) = \chi_n(\xi + \xi_0)\). 
  \end{cor}
  
  \textbf{Remark:} Notice that the wavefunctions of the original problem are obtained from \(\phi(\xi)\) via (\ref{linearderiremoval}). 
   
\section{Conclusions}\label{concsec}
In this note, we revisited the nonlinear but isochronous Li\'enard oscillator (\ref{2}) and discussed some of the classical and quantum aspects. In particular, we demonstrated that generic Li\'enard-type systems can be obtained from a certain family of Levinson-Smith-type equations and in this context, we utilized nonlocal transformations applied to the Levinson-Smith-type equations to demonstrate that the system of our interest supports isochronous oscillations. The Jacobi last multiplier leads to a Lagrangian description should the Chiellini integrability condition be satisfied. The latter implies the existence of two classes of Hamiltonians for the Li\'enard system (\ref{2}). Such Hamiltonians admit the notion of momentum-dependent mass and exhibit interchanged roles of the coordinate and momentum variables (see also, the generalizations presented in Appendix (\ref{app0})). \\

While the classical Hamilton's equations are equivalent for both the Hamiltonians (when they are coupled to give rise to the second-order equation for the coordinate variable), we demonstrated that their quantum-mechanical features are largely different. One of the Hamiltonians which we denoted earlier to be of class-II admits an infinite tower of equispaced energy levels, giving us a nontrivial example of the correspondence between isochronicity and equispaced quantum spectra \cite{isospec}. Further, we showed that the class-I scenario admits branching, leading to a pair of branched Hamiltonians, both of which exhibit peculiarities with the boundary conditions. \\

Li\'enard systems admitting momentum-dependent masses offer an intriguing testing ground for exploring quantum mechanics in the momentum representation. This opens up the possibility of having a relook at such systems from a supersymmetric viewpoint \cite{susy1,susy2,susy3,ortiz}, but in the momentum representation as it is the naturally-adapted setting for these systems. While quantum mechanics in the position representation with position-dependent masses has been explored in some detail in past papers \cite{vonroos,L2quant,L2quant1,VR0,VR1,VR2,VR3,VR4,VR5,bagchi2,mus1,carinena,cruz,que,koc,cunha,mus2,dha,fer,bag5,bpm,susy3}, the momentum-space counterpart remains largely unexplored. To this end, it should be pointed out that one may consider a generalization of (\ref{2}) which follows from the choices
\begin{equation}\label{4}
    f(x) = 3\kappa x + b, \quad \quad g(x) = \omega^2 x + b\kappa x^2 + \kappa^2 x^3,
\end{equation}
involving $b$, $\kappa$, and $\omega$ as arbitrary (real) constants. Notice that the choice \(b = 0\) and \(\kappa = k/3\) gives us (\ref{2}). Such a system admits an eight-parameter Lie group of point transformations and it follows that there exists a complex point transformation mapping (\ref{1}) along with (\ref{4}) to a free-particle system. In \cite{senti3}, such a point transformation was set up and the consequences were examined, including determining the general solution in terms of the initial data. It is straightforward to check that the Li\'enard system with the functions \(f(x)\) and \(g(x)\) as given by (\ref{4}) satisfies the Chiellini integrability condition (\ref{10}) for \(\omega^2 = \frac{2}{9} b^2\). The corresponding Lagrangian and Hamiltonian functions are closely related to those of the case (\ref{2}) under a shift of the coordinate variable. We hope that the present work will initiate new developments on the quantum mechanics of momentum-dependent-mass systems, in particular, concerning their supersymmetric factorization and intertwining relations, following related developments in the position-dependent-mass scenario \cite{susy3}.

\section*{Acknowledgements} 
A.G. would like to thank Miloslav Znojil for insightful comments and gratefully acknowledges the Czech Technical University in Prague for hospitality, where many of the ideas presented in this work were conceived. A.G. also thanks Akash Sinha for various related discussions. A.G.C. and P.G. express their gratitude to Pepin Cari\~nena for stimulating discussions. We would like to thank Jasleen Kaur for help in preparing the figures and for carefully reading the manuscript. A.G. expresses his gratitude to the Ministry of Education, Government of India for financial support in the form of a Prime Minister's Research Fellowship (ID: 1200454).

\appendix

\counterwithin*{equation}{section}
\renewcommand\theequation{\thesection\arabic{equation}}

\section{Generalized class of Li\'enard oscillators with momentum-dependent masses}\label{app0}
A possible generalization of (\ref{11}) would be a situation where
\begin{equation}
\ell = - \frac{1}{r}, \frac{1-r}{r},
\end{equation} such that \(r=3\) gives us (\ref{11}) as a special case. The two values of \(\ell\) add up to \(-1\). Let us take \(f(x) = a_1 x^s\), where \(s \in \mathbb{R}\). The Chiellini condition (\ref{10}) demands that
\begin{equation}
\frac{d}{dx} \big(x^{-s} g(x) \big) = \bigg( \frac{r-1}{r^2} \bigg)a_1^2 x^s.
\end{equation}
This implies that \(g(x)\) must be given by \cite{bag3,pgcar1}
\begin{equation}
g(x) = \bigg( \frac{r-1}{r^2} \bigg) \bigg(\frac{a_1^2}{s+1}\bigg) x^{2s + 1} + a_2 x^s,
\end{equation} where \(a_2\) is the integration constant. Since from (\ref{17}), the appearance of momentum-dependent mass (proportional to the reciprocal of the coefficient of the \(x^2\) term in the Hamiltonian) requires \(g(x)/f(x)\) to be a quadratic, let us put \(s=1\). We thus get a generalized Li\'enard system 
\begin{equation}\label{lienardgeneralche}
\ddot{x} + (a_1x) \dot{x} + \bigg[ \bigg( \frac{r-1}{r^2} \bigg) \bigg(\frac{a_1^2}{2}\bigg) x^{3} + a_2 x \bigg] = 0.
\end{equation}
The system mentioned above is a generalization of (\ref{2}) which admits the notion of momentum-dependent mass. For the class-I and class-II cases corresponding to \(\ell = -1/r\) and \((1-r)/r\), respectively, the Hamiltonians (\ref{17}) turn out to be (ignoring the possibility of branching but which could be trivially taken into account)
\begin{equation}
H(x,\tilde{p}) = \frac{x^2}{2m(\tilde{p})} + V(\tilde{p}), 
\end{equation} where
\begin{equation}
m_{\rm I}(\tilde{p}) = \frac{r}{a_1 \tilde{p}}, \quad \quad  m_{\rm II}(\tilde{p}) = \frac{r}{(r-1)a_1 \tilde{p}},
\end{equation}
\begin{equation}
V_{\rm I}(\tilde{p}) = \bigg( \frac{a_2 r}{a_1 (r-1)} \bigg)\tilde{p} + \frac{\tilde{p}^{\frac{r-2}{r-1}}}{2-r}, \quad \quad V_{\rm II}(\tilde{p}) = \bigg(\frac{ r a_2}{a_1}\bigg) \tilde{p} + \bigg( \frac{1-r}{2-r} \bigg) \tilde{p}^{2-r}. 
\end{equation}
The expressions for the momentum-dependent masses and potentials presented above reduce to those given in (\ref{massprofiles}) and (\ref{potprofiles}) for \(r=3\); in that case, the class-I potentials admit branching due to the \(\sqrt{\tilde{p}}\) term.  However, a straightforward calculation along the lines of Sec. (\ref{levinsonsec}) reveals that unless \(r = 3\), the Li\'enard system (\ref{lienardgeneralche}) does not support isochronous oscillations \cite{sab,isolie}. Thus, although (\ref{lienardgeneralche}) is a generalized family of Li\'enard oscillators admitting the notion of momentum-dependent mass, the case (\ref{2}) investigated in this paper is special. 

\section{Isotonic oscillator}\label{appA}
The spectrum of the equation \cite{isotonic,isotonic1,isotonic2} (\(z > 0\))
\begin{equation}
- \frac{d^2 \Theta(z)}{dz^2} + V(z) \Theta(z) = \mathcal{E} \Theta(z), \quad \quad V(z) = V_0 \bigg(\frac{z}{z_0} - \frac{z_0}{z}\bigg)^2, \quad \quad V_0, z_0>0,
\end{equation} turns out to be
\begin{equation}
\mathcal{E}_n = \frac{4 \sqrt{V_0}}{z_0} \bigg( n + \frac{1}{2} + \frac{1}{4} \Big(\sqrt{1 + 4V_0 z_0^2} - 2 z_0 \sqrt{V_0} \Big) \bigg), \quad \quad n=0,1,2,\cdots.
\end{equation} 
The corresponding solutions read
\begin{equation}
\Theta_n(z) = \lambda_n \bigg(\frac{z}{z_0}\bigg)^\nu e^{-\frac{ K z^2}{4z_0^2}} {_1F_1}\bigg(-n; \nu + \frac{1}{2}; \frac{K z^2}{2 z_0^2} \bigg),
\end{equation} where
\begin{equation}
K^2 = 4 V_0 z_0^2, \quad \quad \nu = \frac{1}{2} \big[\sqrt{1 + K^2} + 1\big],
\end{equation} \({_1F_1}(\cdot;\cdot;\cdot)\) is the confluent hypergeometric function, and \(\{\lambda_n\}\) are normalization constants. The above results can be readily verified and the derivation can be found in \cite{isotonic,isotonic2}.

\end{document}